\documentclass[12pt]{article}

\usepackage{graphicx}

\advance \topmargin by -\headheight
\advance \topmargin by -\headsep
     
\evensidemargin \oddsidemargin
\begin{document}

\topmargin -.6in
\def\nonu{\nonumber}
\def\rf#1{(\ref{eq:#1})}
\def\lab#1{\label{eq:#1}} 
\def\br{\begin{eqnarray}}
\def\er{\end{eqnarray}}
\def\be{\begin{equation}}
\def\ee{\end{equation}}
\def\0{\nonumber}
\def\lb{\lbrack}
\def\rb{\rbrack}
\def\({\left(}
\def\){\right)}
\def\v{\vert}
\def\bv{\bigm\vert}
\def\lskip{\vskip\baselineskip\vskip-\parskip\noindent}
\relax
\newcommand{\nit}{\noindent}
\newcommand{\ct}[1]{\cite{#1}}
\newcommand{\bi}[1]{\bibitem{#1}}
\def\a{\alpha}
\def\b{\beta}
\def\ca{{\cal A}}
\def\cm{{\cal M}}
\def\cn{{\cal N}}
\def\cf{{\cal F}}
\def\d{\delta} 
\def\D{\Delta}
\def\eps{\epsilon}
\def\g{\gamma}
\def\G{\Gamma}
\def\grad{\nabla}
\def\h{ {1\over 2}  }
\def\hc{\hat{c}}
\def\hd{\hat{d}}
\def\hg{\hat{g}}
\def\hp{ {+{1\over 2}}  }
\def\hm{ {-{1\over 2}}  }
\def\k{\kappa}
\def\l{\lambda}
\def\L{\Lambda}
\def\lg{\langle}
\def\m{\mu}
\def\n{\nu}
\def\o{\over}
\def\om{\omega}
\def\O{\Omega}
\def\p{\phi}
\def\pa{\partial}
\def\pr{\prime}
\def\ra{\rightarrow}
\def\rh{\rho}
\def\rg{\rangle}
\def\s{\sigma}
\def\t{\tau}
\def\th{\theta}
\def\ti{\tilde}
\def\wti{\widetilde}
\def\inte{\int dx }
\def\xb{\bar{x}}
\def\yb{\bar{y}}

\def\tr{\mathop{\rm tr}}
\def\Tr{\mathop{\rm Tr}}
\def\partder#1#2{{\partial #1\over\partial #2}}
\def\ds{{\cal D}_s}
\def\wtwo{{\wti W}_2}
\def\lie{{\cal G}}
\def\alie{{\widehat \lie}}
\def\dlie{{\cal G}^{\ast}}
\def\elie{{\widetilde \lie}}
\def\edlie{{\elie}^{\ast}}
\def\hlie{{\cal H}}
\def\wlie{{\widetilde \lie}}

\def\rlx{\relax\leavevmode}
\def\inbar{\vrule height1.5ex width.4pt depth0pt}
\def\IZ{\rlx\hbox{\sf Z\kern-.4em Z}}
\def\IR{\rlx\hbox{\rm I\kern-.18em R}}
\def\IC{\rlx\hbox{\,$\inbar\kern-.3em{\rm C}$}}
\def\one{\hbox{{1}\kern-.25em\hbox{l}}}

\def\PRL#1#2#3{{\sl Phys. Rev. Lett.} {\bf#1} (#2) #3}
\def\NPB#1#2#3{{\sl Nucl. Phys.} {\bf B#1} (#2) #3}
\def\NPBFS#1#2#3#4{{\sl Nucl. Phys.} {\bf B#2} [FS#1] (#3) #4}
\def\CMP#1#2#3{{\sl Commun. Math. Phys.} {\bf #1} (#2) #3}
\def\PRA#1#2#3{{\sl Phys. Rev.} {\bf A#1} (#2) #3}
\def\PRD#1#2#3{{\sl Phys. Rev.} {\bf D#1} (#2) #3}
\def\PRE#1#2#3{{\sl Phys. Rev.} {\bf E#1} (#2) #3}
\def\PLA#1#2#3{{\sl Phys. Lett.} {\bf #1A} (#2) #3}
\def\PLB#1#2#3{{\sl Phys. Lett.} {\bf #1B} (#2) #3}
\def\JMP#1#2#3{{\sl J. Math. Phys.} {\bf #1} (#2) #3}
\def\PTP#1#2#3{{\sl Prog. Theor. Phys.} {\bf #1} (#2) #3}
\def\SPTP#1#2#3{{\sl Suppl. Prog. Theor. Phys.} {\bf #1} (#2) #3}
\def\AoP#1#2#3{{\sl Ann. of Phys.} {\bf #1} (#2) #3}
\def\PNAS#1#2#3{{\sl Proc. Natl. Acad. Sci. USA} {\bf #1} (#2) #3}
\def\RMP#1#2#3{{\sl Rev. Mod. Phys.} {\bf #1} (#2) #3}
\def\PR#1#2#3{{\sl Phys. Reports} {\bf #1} (#2) #3}
\def\AoM#1#2#3{{\sl Ann. of Math.} {\bf #1} (#2) #3}
\def\UMN#1#2#3{{\sl Usp. Mat. Nauk} {\bf #1} (#2) #3}
\def\FAP#1#2#3{{\sl Funkt. Anal. Prilozheniya} {\bf #1} (#2) #3}
\def\FAaIA#1#2#3{{\sl Functional Analysis and Its Application} {\bf #1} (#2)
#3}
\def\BAMS#1#2#3{{\sl Bull. Am. Math. Soc.} {\bf #1} (#2) #3}
\def\TAMS#1#2#3{{\sl Trans. Am. Math. Soc.} {\bf #1} (#2) #3}
\def\InvM#1#2#3{{\sl Invent. Math.} {\bf #1} (#2) #3}
\def\LMP#1#2#3{{\sl Letters in Math. Phys.} {\bf #1} (#2) #3}
\def\IJMPA#1#2#3{{\sl Int. J. Mod. Phys.} {\bf A#1} (#2) #3}
\def\AdM#1#2#3{{\sl Advances in Math.} {\bf #1} (#2) #3}
\def\RMaP#1#2#3{{\sl Reports on Math. Phys.} {\bf #1} (#2) #3}
\def\IJM#1#2#3{{\sl Ill. J. Math.} {\bf #1} (#2) #3}
\def\APP#1#2#3{{\sl Acta Phys. Polon.} {\bf #1} (#2) #3}
\def\TMP#1#2#3{{\sl Theor. Mat. Phys.} {\bf #1} (#2) #3}
\def\JPA#1#2#3{{\sl J. Physics} {\bf A#1} (#2) #3}
\def\JSM#1#2#3{{\sl J. Soviet Math.} {\bf #1} (#2) #3}
\def\MPLA#1#2#3{{\sl Mod. Phys. Lett.} {\bf A#1} (#2) #3}
\def\JETP#1#2#3{{\sl Sov. Phys. JETP} {\bf #1} (#2) #3}
\def\JETPL#1#2#3{{\sl  Sov. Phys. JETP Lett.} {\bf #1} (#2) #3}
\def\PHSA#1#2#3{{\sl Physica} {\bf A#1} (#2) #3}
\def\PHSD#1#2#3{{\sl Physica} {\bf D#1} (#2) #3}


\begin{center}
{\large\bf Nonautonomous mixed mKdV-sinh-Gordon hierarchy}
\end{center}
\normalsize
\vskip .4in

\begin{center}
J.F. Gomes, G. R. de Melo, L.H. Ymai and A.H. Zimerman 

\par \vskip .1in \noindent
Instituto de F\'{\i}sica Te\'{o}rica-UNESP\\
Rua Dr Bento Teobaldo Ferraz 271, Bloco II, \\
01140-070, S\~ao Paulo, Brazil\\
\par \vskip .3in

\end{center}
\baselineskip=1cm
\begin{abstract}

The construction of a nonautonomous mixed mKdV/sine-Gordon  model is proposed  by employing an infinite  dimensional affine Lie algebraic structure within the zero curvature representation.  A systematic construction of soliton solutions 
is provided by an adaptation of the dressing method  which takes into account arbitrary time dependent functions.  A particular choice of those arbitrary functions provides an interesting solution describing the transition of a pure mKdV system  into a pure sine-Gordon soliton.

\end{abstract}

\section{Introduction}

Sometime ago, the study of nonlinear effects in  lattice dynamics under the influence of a weak dislocation potential  has lead to a mixed 
 mKdV/sine-Gordon  equation \cite{konno}.  The system  was shown to admit multisoliton solutions  and an infinite set of conservation laws \cite{konno}.  More recently the two-breather solution was discussed in \cite{leblond2009} in connection with 
 the propagation of few cycle pulses (FCP) in non linear optical media. 
      According to ref. \cite{leblond2009} the general mKdV/sine-Gordon equation, in fact, describes the propagation  of a ultrashort optical pulses in a Kerr media.   Moreover, it was shown in \cite{leblond-sanchez} that,
 when the  ressonance frequency 
of atoms in the physical system  are well above or well below the characteristic  duration of the pulse,  the propagation is described by  the mKdV or sine-Gordon equations respectively.  The main object of this paper is to provide a systematic construction of soliton solutions that describe the {\it transition between the two regimes}, i.e. governed by  the mKdV and sine-Gordon equations.  This is accomplished by considering the mixed integrable model proposed in \cite{konno} with two arbitrary time-dependent coefficients.  In this paper we show the integrability of the mixed model with time dependent coefficients and  that, by suitable choice of these coefficients  as a smooth  step-type functions (as shown in figs. 1 and 3) we obtain  exact solutions for the mKdV-SG transition and hence a more realistic description of such phenomena.

\section{Algebraic Formalism}

In ref. \cite{gmz} the algebraic structure of the mixed  mKdV/sine-Gordon equation    was    formulated 
within the zero curvature representation and  a graded infinite dimensional Lie algebraic structure as we shall now briefly review.  Consider the    associated $\lie = sl(2)$ Lie algebra with  generators 
satisfying $\lb h, E_{\pm \a} \rb = \pm 2E_{\pm \a}, \;
\lb E_{\a}, E_{-\a} \rb= h$ and    grading operator $Q = 2\l{{d} \o {d\l}} + {1 \o 2} h$.  $Q$   decomposes the associated  
affine Lie algebra $\hat{sl}(2)$ into graded subspaces, 
$\hat\lie = \oplus_{i}\lie_i$,  
\be
\lie_{2m} = \{  \l^m h\}, \qquad 
\lie_{2m+1} = \{E_+^{(2m+1)} \equiv \l^m\(E_{\a} + \l E_{-\a}\), \; E_-^{(2m+1)} \equiv \l^m\(E_{\a} - \l E_{-\a}\)\},
\label{1}
\ee
$m=0,\pm1,\pm2,\ldots$.
In  \cite{gmz},  a simple proof that a mixed mKdV/sine-Gordon hierarchy is indeed an integrable model follows from 
the  zero curvature 
representation  of the integrable hierarchy  generated by
\be
\lb \pa_x + E^{(1)}_+ + A_0, \pa_{t_N} + D^{(N)} + D^{(N-1)} + \cdots
D^{(0)}+ D^{(-1)}  \rb = 0,
\label{1}
\ee
 where $D^{(j)} \in \lie_j$ and $ A_0 = v h$  
contains the field  variable $v=v\(x,t\)$. According to the subspace decomposition (\ref{1}) for $N=3$ and $t=t_3$ which corresponds to the mixed mKdV-SG equation. Let us  parametrize, 
\br
D^{(3)}&=& a_3 E^{(3)}_+  + b_3E_-^{(3)}, \qquad    D^{(2)}= c_2\l h, \nonu \\
D^{(1)} &=& a_1 E^{(1)}_+ + b_1E_-^{(1)}, \qquad    D^{(0)}= c_0 h, \nonu \\
D^{(-1)} &=& a_{-1} E^{(-1)}_+ + b_{-1}E_-^{(-1)}.
\label{2}
\er
The grade by grade decomposing   of eqn. (\ref{1}) leads to
\br
&&b_{3}=0,\qquad \partial_x a_{3}=0, \qquad c_{2}=a_{3} v,\qquad b_{1}=\frac{1}{2}\partial_{x}c_{2},\label{eq1} \\
&&\partial_{x} a_{1}+2vb_{1}=0,\qquad \qquad \quad \partial_{x}b_{1}+2va_{1}-2c_{0}=0,\label{eq2} \\
&&\partial_{x} a_{-1}+2vb_{-1}=0,\qquad \qquad \partial_{x} b_{-1}+2va_{-1}=0, 
\label{eq3}
\er
together with the equation of motion
\br
 \partial_xc_0-\partial_tv-2b_{-1}=0.
\label{4}
\er
In solving eqns. (\ref{eq1}) we find
\begin{eqnarray}
a_{3}= a_{3}(t), \qquad b_{1}= \frac{a_{3}(t)}{2} v_{x}, \qquad c_{2}=a_{3}(t) v,\label{eq4}
\end{eqnarray}
where $a_3(t)$ is an arbitrary function of $t$.
Introducing (\ref{eq4}) in the first eqn.  (\ref{eq2}), we obtain
\begin{eqnarray}
\partial_x\left(a_1+a_3(t)\frac{v^2}{2}\right)=0, \nonumber
\end{eqnarray}
which implies that
\begin{eqnarray}
a_1+a_3(t)\frac{v^2}{2}=f_1(t),\nonumber
\end{eqnarray}
where $f_1(t)$ is another arbitrary function of $t$. It therefore follows that
\begin{eqnarray}
a_1=f_1(t)-a_3(t)\frac{v^2}{2}.\label{eq5}
\end{eqnarray}
Substituting (\ref{eq5}) in the second eqn.  (\ref{eq2}), we get 
\begin{eqnarray}
c_0=\frac{a_3(t)}{4}\left(v_{xx}-2v^3\right)+f_1(t)v.\label{c0}
\end{eqnarray}
Adding and subtracting eqns.  (\ref{eq3}), we obtain
\begin{eqnarray}
\partial_x a_\pm=\mp2v a_\pm,\label{eq6}
\end{eqnarray}
where we have denoted
\begin{eqnarray}
a_\pm=a_{-1}\pm b_{-1}.\nonumber
\end{eqnarray}
Without loss of generality we may solve (\ref{eq6}) by changing the variable
\begin{eqnarray}
vh =-\pa_x B B^{-1} = \phi_x h,  \qquad B= e^{-\phi h}, \label{phix}
\end{eqnarray}
which leads us to $a_\pm=f_{-1}(t)e^{\mp2\phi}$,
where $f_{-1}(t)$ is another  arbitrary function of  $t$.
Writing 
\begin{eqnarray}
a_{-1}=\frac{a_++a_-}{2}, \qquad b_{-1}=\frac{a_+-a_-}{2},\nonumber
\end{eqnarray}
we find
\begin{eqnarray}
a_{-1}= f_{-1}(t) \cosh\left( 2 \phi \right), \qquad
b_{-1}= - f_{-1}(t) \sinh\left( 2 \phi \right).\label{b(-1)}
\end{eqnarray}
Substituting (\ref{c0}), (\ref{phix}) and (\ref{b(-1)}) in (\ref{4}), we finally obtain
\begin{equation}
\frac{a_{3}(t)}{4} \left( \phi_{x x x x} - 6 \phi_{x}^{2}
\phi_{x x} \right)+f_1(t)\phi_{xx} -\phi_{x t} + 2 f_{-1}(t) \sinh\left( 2
\phi \right)=0.\label{mkdv_ShG}
\end{equation}

Considering $f_1(t)=0$, $a_3(t) = constant$ and $f_{-1}(t) = constant$ we find the usual mixed mKdV/sine-Gordon equation.  For $f_1(t)=0$, $a_3(t)$ a given numerical constant $\neq 0$  we recover eqn. (10) of ref. \cite{kundu-2009}. 
 Moreover for $f_{-1}(t) =0$, we recover equation considered in \cite{pradan}
with a choice of coefficients  that makes the model  integrable.

We should point out that by  change of coordinates (see for instance \cite{kundu-E}) $(x,t) \rightarrow (\tilde x, \tilde t) = (x+V(t), t)$ where $V_t = f_1(t)$  followed by  a subsequently change $\tilde t \rightarrow T = \int a_3(\tilde t) d \tilde t$ and  re-scaling $f_{-1} \rightarrow \tilde \eta$ leads to 
\begin{equation}
 \frac{1}{4}(\phi_{x x x x} - 6 \phi_{x}^{2}
\phi_{x x})  -\phi_{x t} + 2 \tilde \eta(t) \sinh\left( 2
\phi \right)=0.\label{mkdv_ShG2}
\end{equation}
Although eqn. (\ref{mkdv_ShG2}) corresponds to the  equation discussed in \cite{kundu-2009} the object of this paper is to consider a class  of solutions that interpolates between the mKdV and sine-Gordon equations.  This is more conveniently accomplished by employing eqn. (\ref {mkdv_ShG}) where  the two arbitrary functions $a_3(t)$ and $f_{-1}(t)$  (with $f_1(t) = 0$)  can be chosen as  step-like limiting functions (Figs. 1 and 3) as we shall see.

\section{Construction of Soliton Solutions}

In order to construct, in a systematic manner, the soliton solutions of the mixed  model let  us now recall some  basic aspects  of the dressing method (see for instance \cite{mira}).   The zero curvature  representation (\ref{1}) implies in a pure gauge configuration, i.e.,  
\br
\pa_x + E + A_0 = \pa_x T T^{-1}, \qquad \pa_t + D^{(3)} + \cdots D^{(-1)} = \pa_t T T^{-1},
\er
In particular, the vacuum  is obtained by setting $\phi_{vac} =0$ {\footnote {For a general  member of the hierarchy evolving according $t=t_{2n+1}$, the vacuun configuration implies  
\br
\pa_x T_0 T_0^{-1} = -E^{(1)}_+, \qquad \pa_{t_{2n+1}} T_0T_0^{-1} = -a_{2n+1}(t) E^{(2n+1)}_+ - f_{-1}(t) E^{(-1)}_+ -\sum_{k=1}^{n}f_{2k-1}(t)E_+^{(2k-1)}.
\nonumber
\er }}
which implies,
\br
\pa_x T_0 T_0^{-1} = -E^{(1)}_+, \qquad \pa_t T_0T_0^{-1} = -a_3(t) E^{(3)}_+ - f_{-1}(t) E^{(-1)}_+ -f_1(t)E_+^{(1)}.
\label{6}
\er
 which after integration yields
 \br
 T_0 = \exp \( -\int^t dt^{\pr}a_3(t^{\pr}) E^{(3)}_+ - \int^t dt^{\pr}f_{-1}(t^{\pr}) E^{(-1)}_+- \int^t dt^{\pr}f_{1}(t^{\pr}) E^{(1)}_+\) \exp (-xE^{(1)}_+),
 \label{7}
 \er
 Following the dressing method explained in  \cite{mira} and employied in \cite{gmz}  we define the tau-functions
 \br
 \tau_n  \equiv \langle\l_n | B |\l_n \rangle = \langle \l_n | T_0 g T_0^{-1} |\l_n\rangle, 
 \label{8}
 \er
 where $\l_n, n=0,1$ are fundamental weights of the  full affine Kac-Moody algebra $\hat{sl}(2)$, $g$ is a constant group element which classifies the soliton solutions and $B$ is a zero grade group element containing the physical fields.  
 In order to ensure heighest weight representations
 we now introduce  central extensions within the affine Lie algebra, characterized by $\hat c$, i.e.,
 \begin{eqnarray}
\lbrack h^{(n)},E_{\pm\alpha}^{(m)}\rbrack &=&\pm 2
E_{\pm\alpha}^{(n+m)},\nonumber\\
\lbrack E_{\alpha}^{(n)},E_{-\alpha}^{(m)}\rbrack &=&h^{(n+m)}+n\delta_{n+m,0}\hat{c},\nonumber\\
\lbrack h^{(n)},h^{(m)}\rbrack &=&2n\delta_{n+m,0}\hat{c},\nonumber
\end{eqnarray}
and  define highest weight representations, i.e.,
\br
h|\l_n\rangle =  \d_{n,1} |\l_n\rangle, \qquad \hat c |\l_n\rangle =  |\l_n\rangle, \qquad \lie_i |\l_n\rangle = 0, \quad i>0,
\er
$n=0,1$. Under this  affine picture the group element $B$ acquires  a central term contribution,
\br
B= e^{-\phi h} e^{-\nu \hat c}.
\er
In order to  obtain explicit space-time dependence from the r.h.s. of (\ref{8}) we consider the vertex operators,
\begin{equation}
V\left( \gamma \right)=\sum_{n=-\infty}^{\infty} ( \l^n h-\frac{1}{2}
\hat{c}\delta_{n,0})\gamma^{-2n}+ E_-^{(2n+1)} \gamma^{-2n-1},\nonumber
\end{equation}
satisfying 
\begin{equation}
\left[ E_+^{(2n+1)} ,V\left( \gamma \right) \right] =-2 \gamma^{2n+1} V\left(
\gamma \right).\nonumber
\end{equation}
For a general $M$-soliton solution   the  group element $g$ in (\ref{8}) is written as
\begin{eqnarray}
g=\prod_{j=1}^{M} e^{\alpha_jV(\gamma_j)},\label{g}
\end{eqnarray}
where $\alpha_j$ are arbitrary constants. 
We therefore obtain
\begin{eqnarray}
\tau_0&=&e^{-\nu}=\langle\lambda_0|\prod_{j=1}^{M}e^{\alpha_j\rho_j(x,t)V(\gamma_j)}|\lambda_0\rangle,\nonumber\\
\tau_1&=&e^{-\phi-\nu}=\langle\lambda_1|\prod_{j=1}^{M}e^{\alpha_j\rho_j(x,t)V(\gamma_j)}|\lambda_1\rangle,\nonumber
\end{eqnarray}
where {\footnote { In considering a general $(2n+1)$-th member of the hierarchy, 
\begin{eqnarray}
\rho_j(x,t)=e^{2\gamma_j x+2\gamma_j^{2n+1}A_{2n+1}(t)+2\sum_{k=1}^n \gamma_j^{2k-1}F_{2k-1}(t)+2\gamma_j^{-1}A_{-1}(t)},\quad A_{2n+1}(t)=\int dt\, a_{2n+1}(t),\quad F_{2k-1}= \int dt\, f_{2k-1}(t). \nonumber
\end{eqnarray}}}
\begin{eqnarray}
\rho_j(x,t)=e^{2\gamma_j x+2\gamma_j^3A_3(t)+2\gamma_jF_1(t)+2\gamma_j^{-1}A_{-1}(t)},\label{rho}
\end{eqnarray}
\begin{eqnarray}
A_3(t)=\int dt\, a_3(t), \qquad F_1(t)=\int dt\, f_1(t), \qquad A_{-1}(t)=\int dt\, f_{-1}(t).\nonumber
\end{eqnarray}
As an illustrative example, we consider the one and  two soliton cases,  $M=1,2$, where
\begin{eqnarray}
   \tau_{0}^{1-sol} = e^{-\nu} = 1 - \frac{\alpha_1}{2} \rho_{1},\qquad
\tau_{1}^{1-sol} = e^{-\nu - \phi } = 1 + \frac{\alpha_1}{2} \rho_{1},
\label{1-sol}
\end{eqnarray}
and
\begin{eqnarray}
   \tau_{0}^{2-sol} &=& e^{-\nu} = 1 - \frac{\alpha_1}{2} \rho_{1}- \frac{\alpha_2}{2} \rho_{2}+ \alpha_1\alpha_2 A_{1,2} \rho_{1} \rho_{2},\nonumber\\
\tau_{1}^{2-sol} &=& e^{-\nu - \phi } = 1 + \frac{\alpha_1}{2} \rho_{1}+ \frac{\alpha_2}{2} \rho_{2}+ \alpha_1\alpha_2 A_{1,2} \rho_{1} \rho_{2},
\label{2-sol}
\end{eqnarray}
respectively.  In order to obtain (\ref{1-sol}) and (\ref{2-sol})
where we have used the fact that
\begin{eqnarray}
\langle\lambda_n|V(\gamma)|\lambda_n\rangle&=&\delta_{n,1}-\frac{1}{2}, \qquad \qquad n=0,1\nonumber\\
\langle\lambda_n|V(\gamma_1)V(\gamma_2)|\lambda_n\rangle&=&A_{1,2}=\frac{1}{4}\left( \frac{\gamma_{1}-\gamma_{2}}{\gamma_{1}+\gamma_{2}} \right)^{2}.\nonumber
\end{eqnarray}
The general two soliton solution can then be written as 
\begin{eqnarray}
\phi=\mathrm{ln}\left(\frac{\tau_0}{\tau_1}\right)=\mathrm{ln}\left(\frac{1 - \frac{\alpha_1}{2} \rho_{1}- \frac{\alpha_2}{2} \rho_{2}+ \alpha_1\alpha_2 A_{1,2} \rho_{1} \rho_{2}}{1 + \frac{\alpha_1}{2} \rho_{1}+ \frac{\alpha_2}{2} \rho_{2}+ \alpha_1\alpha_2 A_{1,2} \rho_{1} \rho_{2}}\right),\label{solution}
\end{eqnarray}
while the one soliton  is obtained from (\ref{solution}) by setting $\a_2=0$.
\section{Applications}

According to ref. \cite{leblond2009} the propagation of a FCP with frequency $ \omega $  on a dielectric media with characteristic frequency $\O$, $\omega  << \O$ can be described  by the mKdV equation 
\begin{eqnarray}
\phi_{z \tau}+a\left(\frac{3}{2}\phi_{\tau}^2\phi_{\tau\tau}+\phi_{\tau\tau\tau\tau}\right)=0,\nonumber
\end{eqnarray}
where the coordinates $z$ and  $\tau$ correspond  respectively to the  propagation distance and retarded time, while the electric field  $E= \phi_{\tau}$.
For the case where $\omega>>\Omega$, the system is described by the sine-Gordon equation, 
\begin{eqnarray}
\phi_{z\tau}-b\sin \phi=0. \nonumber
\end{eqnarray}
If we now consider a dielectric media  with two  characteristic frequencies $\Omega_1, \Omega_2$,  the case in the regime $\Omega_1<<\omega<<\Omega_2$,
is described  by the mixed mKdV-SG equation ,
\begin{eqnarray}
\phi_{z \tau}+a\left(\frac{3}{2}\phi_{\tau}^2\phi_{\tau\tau}+\phi_{\tau\tau\tau\tau}\right)-b\sin \phi=0,\nonumber
\end{eqnarray}
where the two constants $a$ and $b$ are related  to the non-linear and dispersion properties of the media.

In order to adapt  model (\ref{mkdv_ShG}) to such situation, define 
\begin{eqnarray}
a_3(t)=-4a\theta_1(t), \qquad f_1(t)=0, \qquad f_{-1}(t)=\frac{b}{4}\theta_2(t),\nonumber
\end{eqnarray}
re-scaling  $\phi\to\frac{i}{2}\phi$, $t\to z$, $x\to\tau$,
eqn.  (\ref{mkdv_ShG}) becomes
\begin{eqnarray}
a\,\theta_1(z)\left(\phi_{\tau\tau\tau\tau}+\frac{3}{2}\phi_{\tau}^2\phi_{\tau\tau}\right)+\phi_{z\tau}-b\,\theta_2(z)\sin\phi=0.\label{equation}
\end{eqnarray}
If we now substitute $
\alpha_k\to -2i\alpha_k, $
and make use of the identity
\begin{eqnarray}
\arctan X=\frac{1}{2i}\mathrm{ln}\left(\frac{1+iX}{1-iX}\right),\nonumber
\end{eqnarray}
we find that the two soliton  solution (\ref{solution}) may be written as 
\begin{eqnarray}
\phi=4\arctan\left(\frac{\alpha_1\rho_1+\alpha_2\rho_2}{1-4\alpha_1\alpha_2A_{1,2}\rho_1\rho_2}\right),\label{solution2}
\end{eqnarray}
where
\begin{eqnarray}
\rho_j=\exp\left(2\gamma_j\tau+2\gamma_j^3A_3(z)+2\gamma_j^{-1}A_{-1}(z)\right),\nonumber
\end{eqnarray}
\begin{eqnarray}
A_3(z)=-4a\int^z dz^{\pr}\,\theta_1(z^{\pr}),\qquad A_{-1}(z)=\frac{b}{4}\int^z dz^{\pr}\,\theta_2(z^{\pr}).\label{theta}
\end{eqnarray}
\subsection{Transition mKdV-SG}

Consider now a dielectric media  in which 
$\Omega_1>>2\gamma_j$ in the region  $z<z_1$ and  $\Omega_2<<2\gamma_j$ in the region  $z>z_2$ with $z_1>z_2$ such that there exist
an overlap region  in which the media admits the two characteristic frequencies, $\Omega_2<< 2\gamma_j <<\Omega_1$.
As an example  to describe such a realistic situation, we take  both $\theta_1(z)$ and  $\theta_2(z)$ as step-like functions (see fig. 1) with 
\begin{eqnarray}
\theta_1(z)=\frac{1}{2}-\frac{1}{\pi}\arctan\left[\beta_1(z-z_1)\right],\qquad
\theta_2(z)=\frac{1}{2}+\frac{1}{\pi}\arctan\left[\beta_2(z-z_2)\right],\nonumber
\end{eqnarray}
where  $\beta_1$ e $\beta_2$ are phenomenological parameters describing the transition between the two medias (see fig. 1).  
\begin{figure}[h]
\center
\includegraphics[width=7cm]{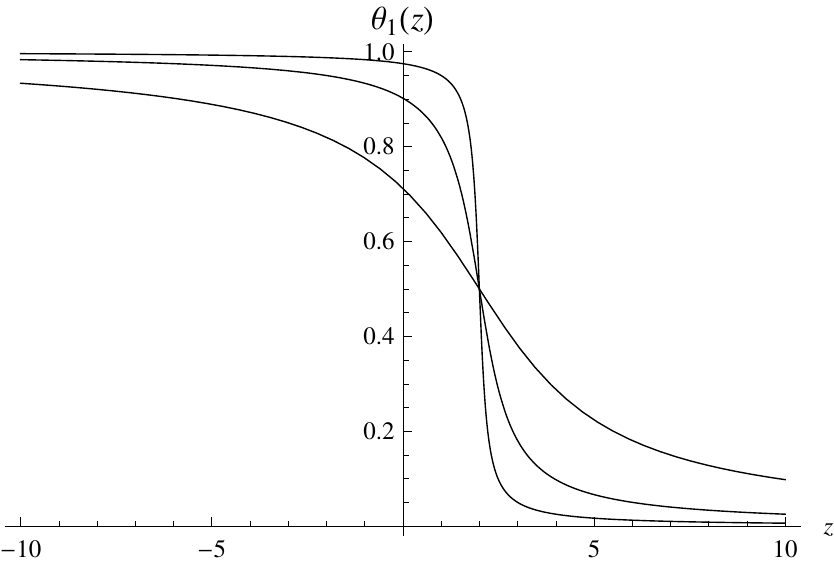}\hspace{1.0cm}
\includegraphics[width=7cm]{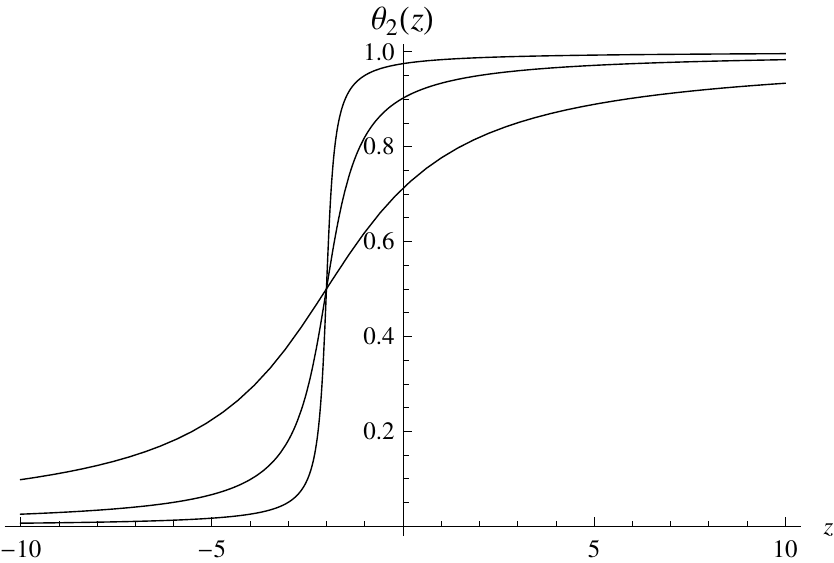}
\caption{Plots of $\theta_1(z)$ and $\theta_2(z)$ for $\beta_1=\beta_2=\{\frac{\pi}{8}, \frac{\pi}{2}, 2\pi$\}, with $z_1=2$ and $z_2=-2$.}
\end{figure}

They can therefore be integrated from (\ref{theta}) to yield,
\begin{eqnarray}
-\frac{1}{4a}A_3(z)&=&\frac{z}{2}-\frac{(z-z_1)}{\pi}\arctan\left[\beta_1(z-z_1)\right]+\frac{1}{2\pi\beta_1}\mathrm{\ln}\left[1+\beta_1^2(z-z_1)^2\right],\nonumber\\
\frac{4}{b}A_{-1}(z)&=&\frac{z}{2}+\frac{(z-z_2)}{\pi}\arctan\left[\beta_2(z-z_2)\right]-\frac{1}{2\pi\beta_2}\mathrm{\ln}\left[1+\beta_2^2(z-z_2)^2\right].\nonumber
\end{eqnarray}
The $\beta_1,\beta_2\to+\infty$ limit, for $z_1=z_2=0$ correspond  to the system governed by the pure mKdV in the  region $z<0$ and by the pure sine-Gordon equation for $z>0$.  In such limit, we have,
\begin{eqnarray}
\theta_1(z)=\frac{1}{2}\left(1-\frac{|z|}{z}\right), \qquad \theta_2(z)=\frac{1}{2}\left(1+\frac{|z|}{z}\right),\label{theta}
\end{eqnarray}
and
\begin{eqnarray}
-\frac{1}{4a}A_3(z)=\frac{1}{2}\left(z-|z|\right), \qquad \frac{4}{b}A_{-1}(z)=\frac{1}{2}\left(z+|z|\right).\label{As}
\end{eqnarray}

Figure 2 below shows the transition mKdV-SG  for the  one soliton solution, $\a_1 = 1, \a_2 =0$.  The plot on the right shows the soliton solution viewed from the above and displays the transition (different velocities) from the mKdV to the SG solitons.
\begin{figure}[h]
\center
\includegraphics[width=6cm]{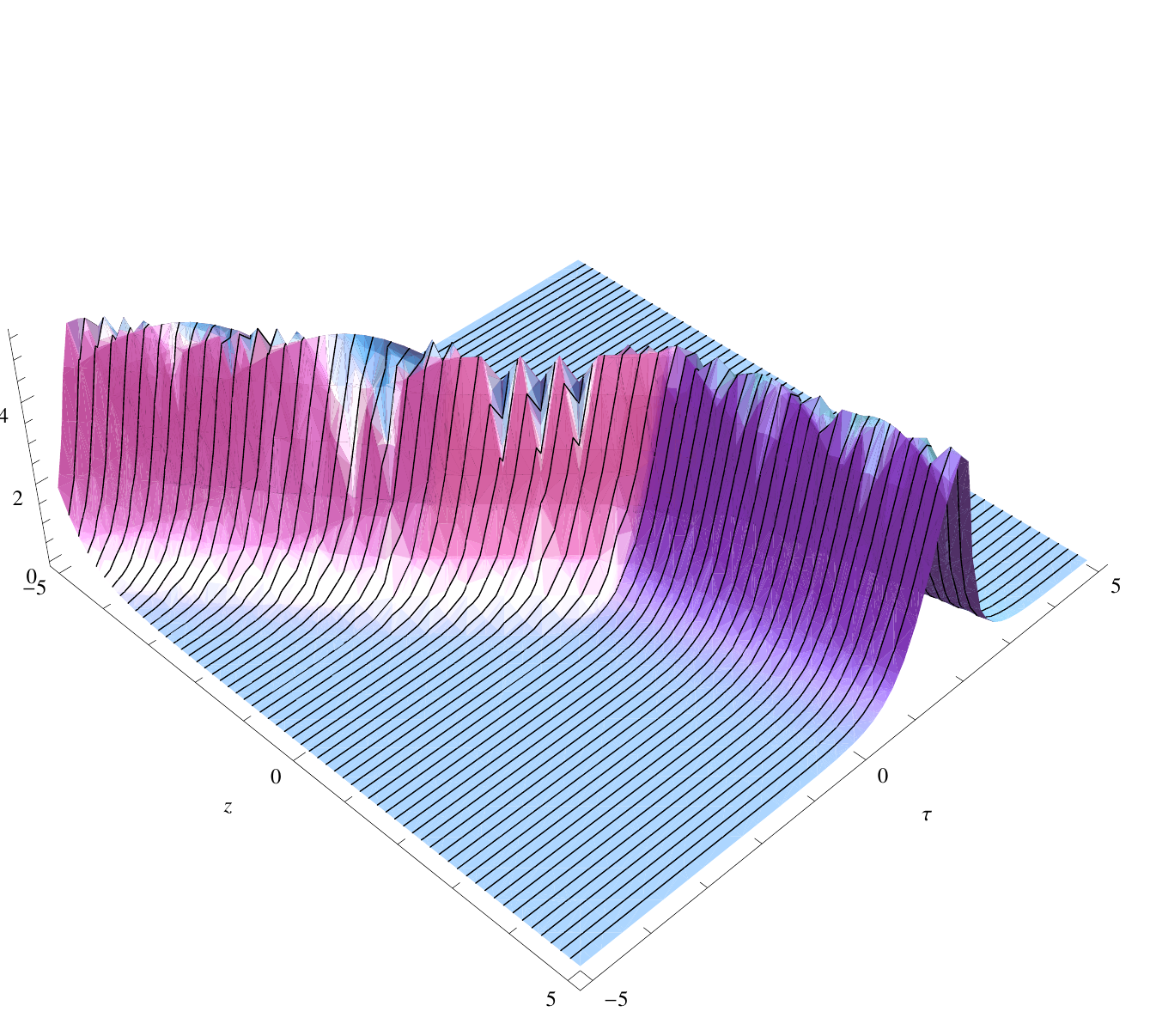}\hspace{1.0cm}
\includegraphics[width=6cm]{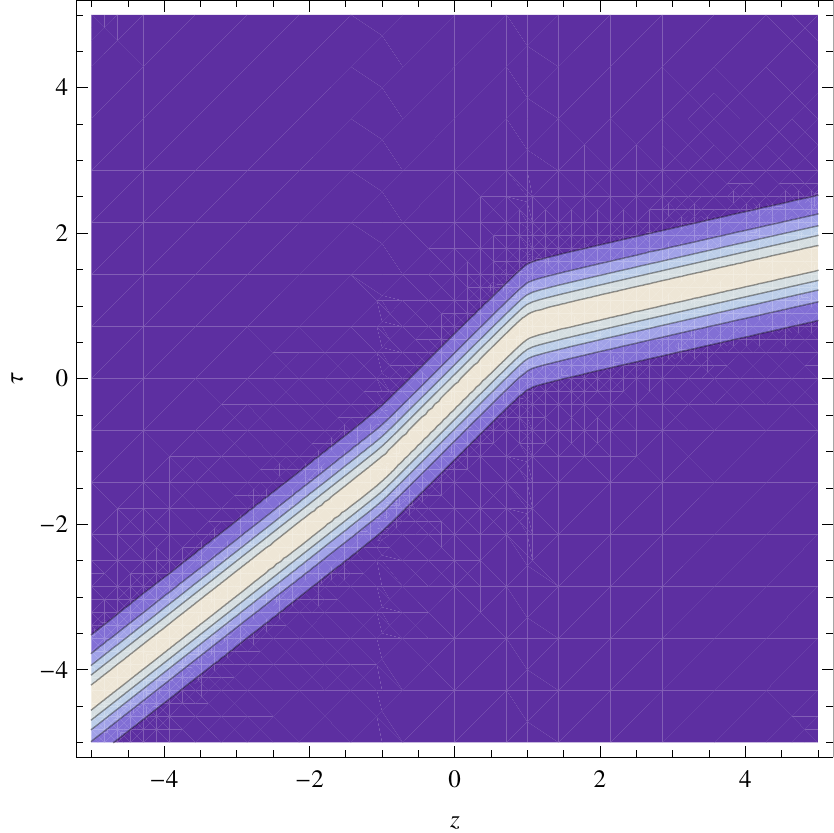} 
\caption{Plot (left) and contour plot (right) of $\phi_\tau$ with $a=\frac{1}{10}$, $b=-\frac{7}{4}$, $\beta_1=\beta_2=8\pi$, $z_1=1$, $z_2=-1$ and $\gamma_1=\frac{14}{10}$.}
\end{figure}  

\subsection{Transition mKdV-SG-mKdV}

Another  example consist in two equal media separated by a second one describing, for instance  the mKdV-SG-mKdV transition.  Mathematically this situation may be described 
by  combining  theta-type functions (see fig. 3), i.e.,
\begin{eqnarray}
\theta_1(z)&=&1-\frac{1}{\pi}\arctan\left[\bar{\beta}_1(z-\bar{z}_1)\right]+\frac{1}{\pi}\arctan\left[\bar{\beta}_2(z-\bar{z}_2)\right],\qquad\bar{z}_1<\bar{z}_2, \nonumber\\
\theta_2(z)&=&\frac{1}{\pi}\arctan\left[\beta_1(z-z_1)\right]-\frac{1}{\pi}\arctan\left[\beta_2(z-z_2)\right],\qquad z_1<z_2,\nonumber
\end{eqnarray} 
with $z_1 \leq \bar z_1, $ and $\bar z_2 \leq z_2$ to guarantee the existence of an overlap region between the two medias.

\begin{figure}[h]
\center
\includegraphics[width=6cm]{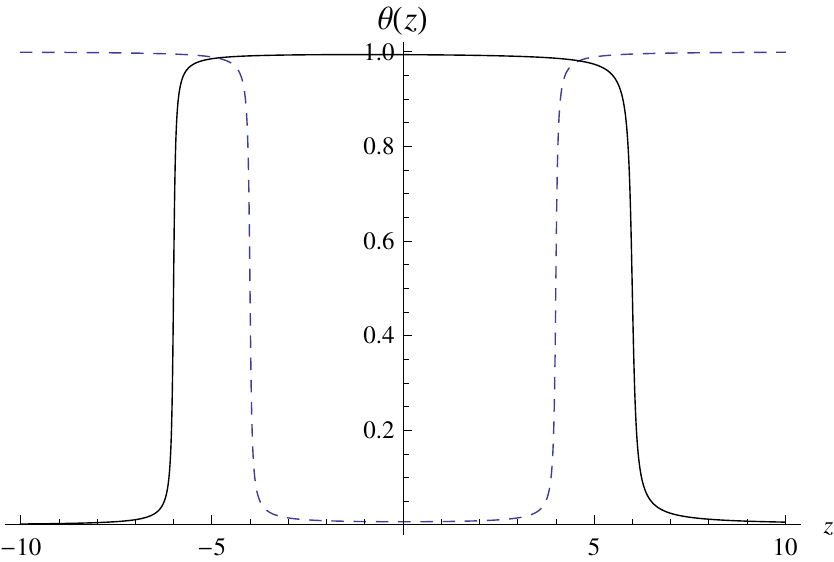}
\caption{Plot of $\theta_1(z)$ (dashed) and $\theta_2(z)$ (continuum) with $\beta_1=\beta_2=\bar{\beta}_1=\bar{\beta}_2=8\pi$, $z_1=-6$, $z_2=6$, $\bar{z}_1=-4$ and $\bar{z}_2=4$.}
\end{figure}

After integration (\ref{theta}) we find
\begin{eqnarray}
-\frac{1}{4a}A_3(z)&=&z-\frac{(z-\bar{z}_1)}{\pi}\arctan\left[\bar{\beta}_1(z-\bar{z}_1)\right]+\frac{(z-\bar{z}_2)}{\pi}\arctan\left[\bar{\beta}_2(z-\bar{z}_2)\right]\nonumber\\
&&+\frac{1}{2\pi\bar{\beta}_1}\ln\left[1+\bar{\beta}_1^2(z-\bar{z}_1)^2\right]-\frac{1}{2\pi\bar{\beta}_2}\ln\left[1+\bar{\beta}_2^2(z-\bar{z}_2)^2\right],\nonumber\\
\frac{4}{b}A_{-1}(z)&=&\frac{(z-z_1)}{\pi}\arctan\left[\beta_1(z-z_1)\right]-\frac{(z-z_2)}{\pi}\arctan\left[\beta_2(z-z_2)\right]\nonumber\\
&&-\frac{1}{2\pi\beta_1}\ln\left[1+\beta_1^2(z-z_1)^2\right]+\frac{1}{2\pi\beta_2}\ln\left[1+\beta_2^2(z-z_2)^2\right],\nonumber
\end{eqnarray}  
The limit  $\beta_1,\beta_2,\bar{\beta}_1,\bar{\beta}_2\to+\infty$, when $z_1=\bar{z}_1$ e $z_2=\bar{z}_2$ with $z_1<z_2$, corresponds to the pure   mKdV case in the region $z<z_1$  and $z>z_2$, and pure 
sine-Gordon in the region $z_1<z<z_2$. Under such limiting case we have
\begin{eqnarray}
\theta_1(z)=1-\frac{1}{2}\left(\frac{|z-z_1|}{(z-z_1)}-\frac{|z-z_2|}{(z-z_2)}\right),\qquad \theta_2(z)=\frac{1}{2}\left(\frac{|z-z_1|}{(z-z_1)}-\frac{|z-z_2|}{(z-z_2)}\right),\nonumber
\end{eqnarray} 
such that
\begin{eqnarray}
-\frac{1}{4a}A_3(z)=z-\frac{1}{2}\left(|z-z_1|-|z-z_2|\right),\qquad \frac{4}{b}A_{-1}(z)=\frac{1}{2}\left(|z-z_1|-|z-z_2|\right).\nonumber
\end{eqnarray}
Figure 4 represents the mKdV-SG-mKdV transition for one soliton solution.
\begin{figure}[h]
\center
\includegraphics[width=6cm]{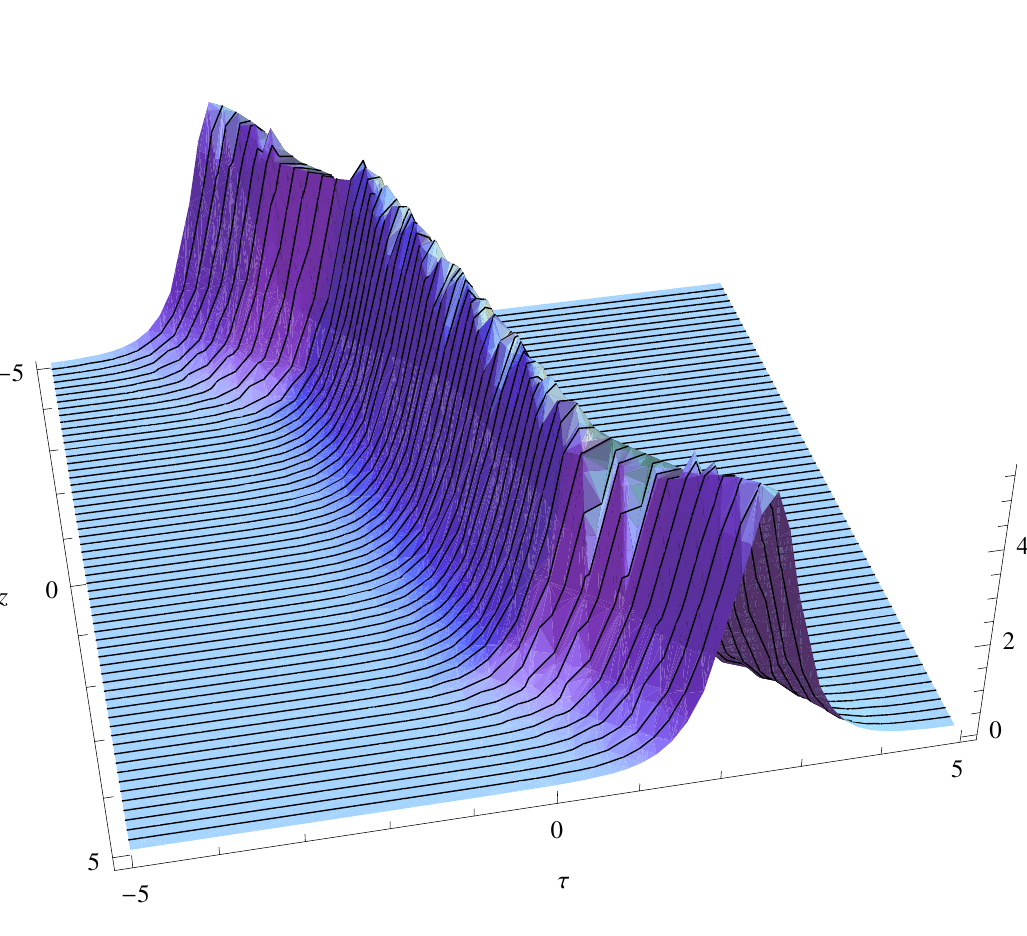}\hspace{1.0cm}
\includegraphics[width=6cm]{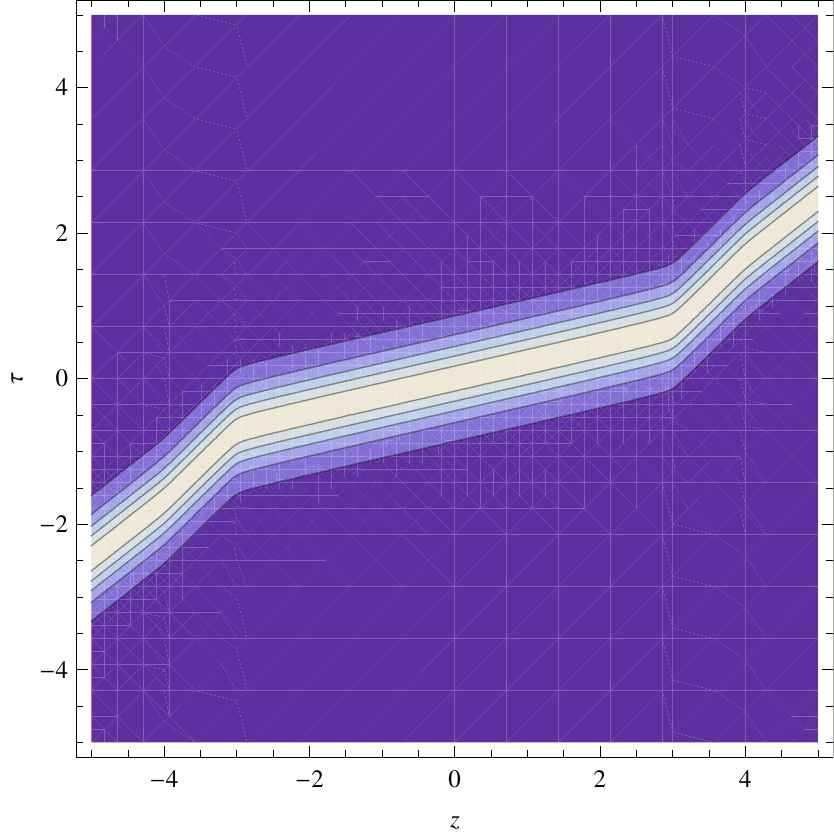} 
\caption{Plot (left) and contour plot (right) of $\phi_\tau$ with $a=\frac{1}{10}$, $b=-\frac{7}{4}$, $\beta_1=\beta_2=\bar{\beta}_1=\bar{\beta}_2=8\pi$, $z_1=-4$, $z_2=4$, $\bar{z}_1=-3$, $\bar{z}_2=3$ and $\gamma_1=\frac{14}{10}$.}
\end{figure} 

\subsection{Two soliton solution}

The two soliton solution ($\a_1 = \a_2 =1$) can be represented by
\begin{figure}[h]
\center
\includegraphics[width=6cm]{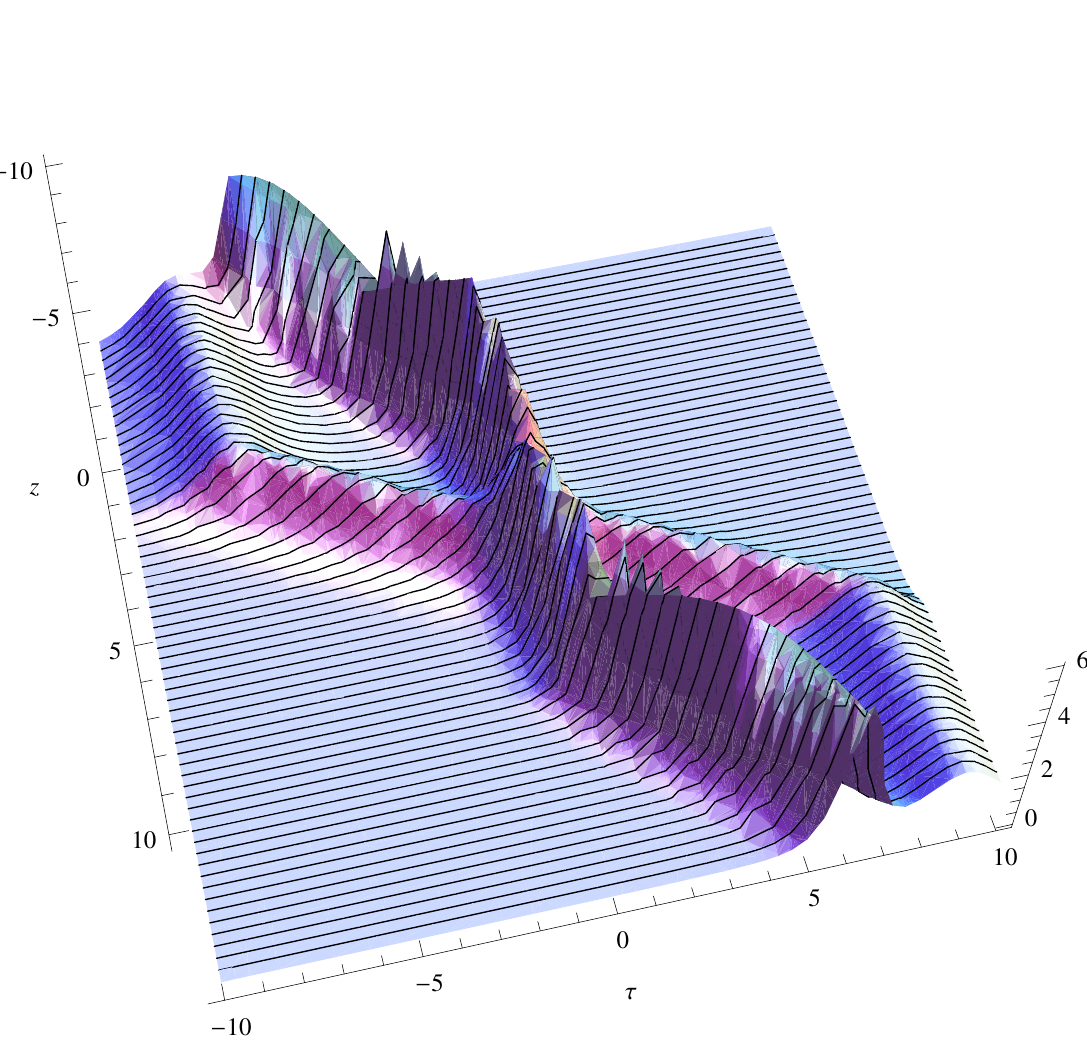}\hspace{1.0cm}
\includegraphics[width=6cm]{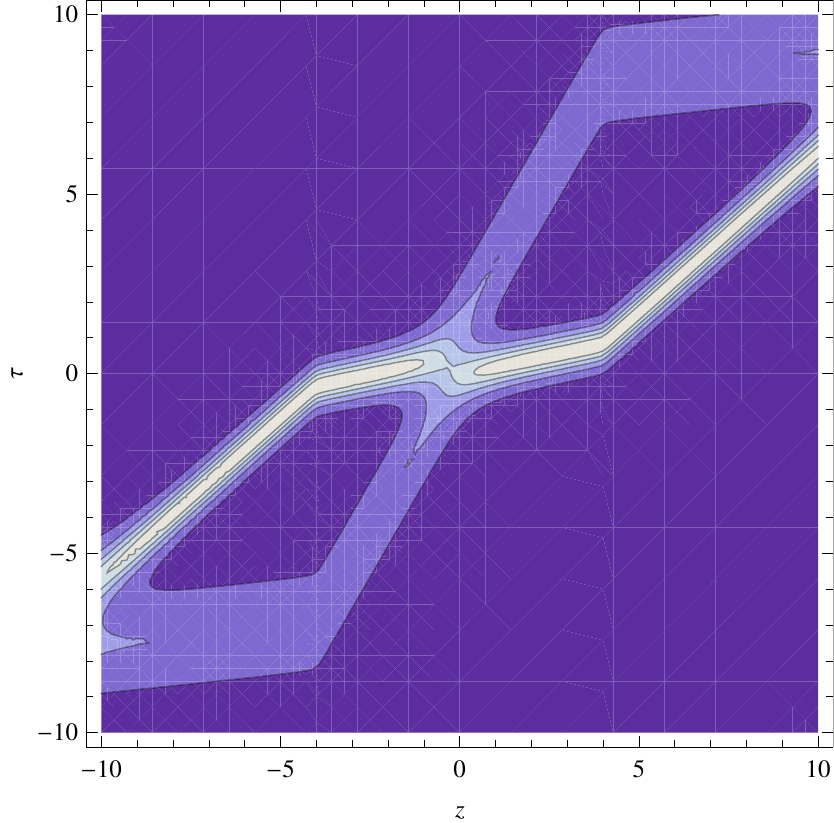} 
\caption{Plot (left) and contour plot (right) of $\phi_\tau$ with $a=\frac{1}{10}$, $b=-\frac{7}{4}$, $\beta_1=\beta_2=\bar{\beta}_1=\bar{\beta}_2=8\pi$, $z_1=\bar{z}_1=-4$, $z_2=\bar{z}_2=4$, $\gamma_1=\frac{3}{2}$ e $\gamma_2=\frac{1}{2}$.}
\end{figure}

As a conclusion, we have  adapted the general dressing construction of soliton solutions to the mixed mKdV-SG hierarchy with arbitrary ``time'' dependent functions.  The choice  of such  arbitrary  functions  as step-type functions allowed exact solutions describing  smooth  transitions  from the mKdV to sine-Gordon regime and therefore more realistic models.  
\vskip .5cm

\newpage
 \noindent
{\bf Acknowledgements} \\
\vskip .05 cm \noindent
{  LHY and GRM acknowledges support from Fapesp and  Capes respectively, JFG and AHZ thank CNPq for partial support.}

\end{document}